\documentclass[%
reprint,
superscriptaddress,
longbibliography,
amsmath,amssymb,
prb,
]{revtex4-2}

\usepackage{graphicx}% Include figure files
\usepackage{dcolumn}% Align table columns on decimal point
\usepackage{bm}% bold math
\usepackage[colorlinks,linkcolor=blue, urlcolor=blue, anchorcolor=blue, citecolor=blue]{hyperref}% add hypertext capabilities
\usepackage{subfigure}
\usepackage{physics}
\usepackage{float}

\begin{document}
\title{Nonreciprocal Relaxation Acceleration}

\author{Xingyu Zhang}
\affiliation{School of Science, Zhejiang University of Science and Technology, Hangzhou 310023, China}

\author{Yihan Ma}
\affiliation{School of Science, Zhejiang University of Science and Technology, Hangzhou 310023, China}

\author{Yue Liu}
\affiliation{Yukawa Institute for Theoretical Physics, Kyoto University, Kyoto 606-8502, Japan}

\author{Niaz Ali Khan}
\affiliation{Department of Physics, Xiamen University, Xiamen 361005, China}

\author{Chenlong Huang}
\affiliation{Department of Physics, Xiamen University, Xiamen 361005, China}

\author{Yuguo Su}\email{suyuguo@zust.edu.cn}
\affiliation{School of Science, Zhejiang University of Science and Technology, Hangzhou 310023, China}

\author{Junyan Luo}
\affiliation{School of Science, Zhejiang University of Science and Technology, Hangzhou 310023, China}

\author{Dahai He}\email{dhe@xmu.edu.cn}
\affiliation{Department of Physics, Xiamen University, Xiamen 361005, China}

\begin{abstract}		
	Driven by recent discoveries regarding the quantum Mpemba effect, the anomalous relaxation dynamics of open quantum systems have garnered significant attention. While expediting thermalization to equilibrium has been extensively studied, dynamically accelerating the convergence toward nonequilibrium steady states remains a formidable challenge. In this article, we find a transient engineered nonreciprocal dissipative channel can provide a
	shortcut that accelerates convergence to the target reciprocal nonequilibrium steady state for the
	considered two-mode model and initial states. Using interacting bosonic modes, we demonstrate that the temporal activation of a nonreciprocal channel efficiently suppresses prolonged inter-mode energy oscillations, enforcing a rapid, unidirectional thermal dump into the environment. Counterintuitively, we find that this relaxation speedup is robust and independent of the direction of the nonreciprocity. Our results provide a powerful thermodynamic technique for rapid state preparation and cooling in continuous-variable quantum systems, particularly critical for low-temperature quantum information processing.
\end{abstract}
\maketitle
\section{introduction}
The quest to comprehend and control the relaxation timescales of open quantum systems has recently experienced a resurgence of interest, largely inspired by quantum analogues of the Mpemba effect \citep{mpemba1969cool,doi:10.1073/pnas.1701264114,ares_quantum_2025} and other anomalous relaxation phenomena \citep{TEZA20261,g94p-7421,dvrt-n2hq,PhysRevLett.134.107101,PhysRevLett.133.010403,zjdz-rqqd,PhysRevLett.131.080402,PhysRevLett.133.010402,PhysRevLett.134.220403,PhysRevLett.133.140404}. Accelerating these dissipative dynamics is not merely of fundamental theoretical interest, it is critically important for optimizing protocols in quantum state preparation \citep{PhysRevLett.134.050603}, quantum error correction \citep{Shtanko2025boundsautonomous}, and the rapid initialization of quantum thermodynamic devices \citep{Pedram_2023}.
However, the majority of existing investigations into accelerated relaxation are restricted to systems thermalizing toward a strict thermal equilibrium. In the broader context of nonequilibrium physics and modern quantum technologies, systems typically relax toward a nonequilibrium steady state (NESS). This is particularly evident in continuous-variable bosonic systems engaged in quantum transport or operating as quantum heat engines, where multiple reservoirs impose sustained thermal gradients \citep{PhysRevE.87.012109,annurev-physchem-040513-103724}. Understanding how to dynamically steer and accelerate the evolution toward a NESS remains a challenging, yet highly desirable, objective.

A defining characteristic of a NESS is the persistent exchange of energy and matter, which manifests as macroscopic non-vanishing currents even after the system has reached its steady state. This inherent transport implies a fundamental directional bias in the system's dynamics. Consequently, to manipulate the relaxation timescale toward such a state, it is physically intuitive to introduce a control mechanism that explicitly breaks the reciprocal symmetry of the underlying interactions.
Nonreciprocity emerges as an ideal and powerful paradigm for this purpose \citep{10.21468/SciPostPhysLectNotes.44,PhysRevLett.131.113602,PhysRevLett.132.210402,PhysRevX.15.011010,PhysRevApplied.22.064072,67wh-1fxv}. 
By enforcing unidirectional energy routing and suppressing detrimental back-action, nonreciprocal interactions have demonstrated profound utility in non-Hermitian physics, directional signal amplification, and quantum control \citep{PhysRevX.5.021025}. Crucially, these theoretical paradigms have recently witnessed rapid experimental translation. State-of-the-art quantum platforms, most notably superconducting microwave circuits \citep{PhysRevX.5.041020} and cavity optomechanical architectures \citep{shen_experimental_2016,fang_generalized_2017}, have successfully implemented robust nonreciprocal devices.
Given that introducing nonreciprocity dynamically alters the spectrum of the system's Liouvillian superoperator, a fundamental question naturally arises: can such nonreciprocal dynamics be harnessed to circumvent conventional reciprocal bottlenecks and explicitly accelerate nonequilibrium relaxation?

In this article, we propose a theoretical scheme harnessing nonreciprocal dissipation to achieve relaxation acceleration toward a NESS. We investigate a prototypical continuous-variable model consisting of two interacting bosonic modes, each coupled to an independent local thermal bath at a distinct temperature. The requisite nonreciprocity is dynamically engineered through the modes' interaction with a shared dissipative reservoir. By analyzing the evolution of both the macroscopic state displacement and the thermodynamic heat fluxes, we demonstrate that a transient, finite-duration activation of the nonreciprocal channel drastically accelerates the convergence to the target NESS. Furthermore, we discuss the practical implementation of this speedup utilizing temporal pulse control, offering a robust strategy for accelerating nonequilibrium quantum transport.

The remainder of this paper is organized as follows. In Sec. \ref{sec2}, we introduce the nonreciprocal two-mode model and theoretically elucidate the origin of the relaxation speedup through the dynamics of the first and second moments (covariance matrix). In Sec. \ref{sec3}, we analyze this phenomenon from a thermodynamic perspective by evaluating the transient heat currents. In Sec. \ref{sec4}, we propose a practical implementation of this nonreciprocal acceleration utilizing temporal pulses. Finally, in Sec. \ref{sec5}, we summarize our main conclusions and outline perspectives for future research.
\section{model and nonreciprocal speedup}\label{sec2}

\subsection{Model}
We begin by considering a standard two-mode nonreciprocal model \cite{PhysRevLett.131.113602}, where
each mode is locally coupled to an independent finite-temperature
reservoir. Assuming the bosonic modes are nonreciprocally coupled
through a shared bath, the time evolution of the reduced density matrix
is governed by the Lindblad master equation

\begin{align}
\frac{d}{dt}\tilde{\rho}= & -i[H_{0},\tilde{\rho}]+\Gamma\mathcal{D}[L]\tilde{\rho}\nonumber \\
 & +\sum_{j=1,2}\kappa_{j}(n_{j}+1)\mathcal{D}[a_{j}]\tilde{\rho}+\kappa_{j}n_{j}\mathcal{D}[a_{j}^{\dagger}]\tilde{\rho},\label{originalH}
\end{align}
where $H_{0}=\omega_{1}a_{1}^{\dagger}a_{1}+\omega_{2}a_{2}^{\dagger}a_{2}+\frac{\lambda}{2}(a_{1}a_{2}^{\dagger}+a_{2}a_{1}^{\dagger})$
is the Hamiltonian of the two-mode bosonic system, and $L=a_{1}-ie^{i\theta}a_{2}$
is the jump operator associated with the nonreciprocal dissipation
strength $\Gamma$. Here, $\mathcal{D}[o]\tilde{\rho}=o\tilde{\rho}o^{\dagger}-\frac{1}{2}\{o^{\dagger}o,\tilde{\rho}\}$
denotes the Lindblad dissipator.
\begin{figure}
\includegraphics[scale=0.4]{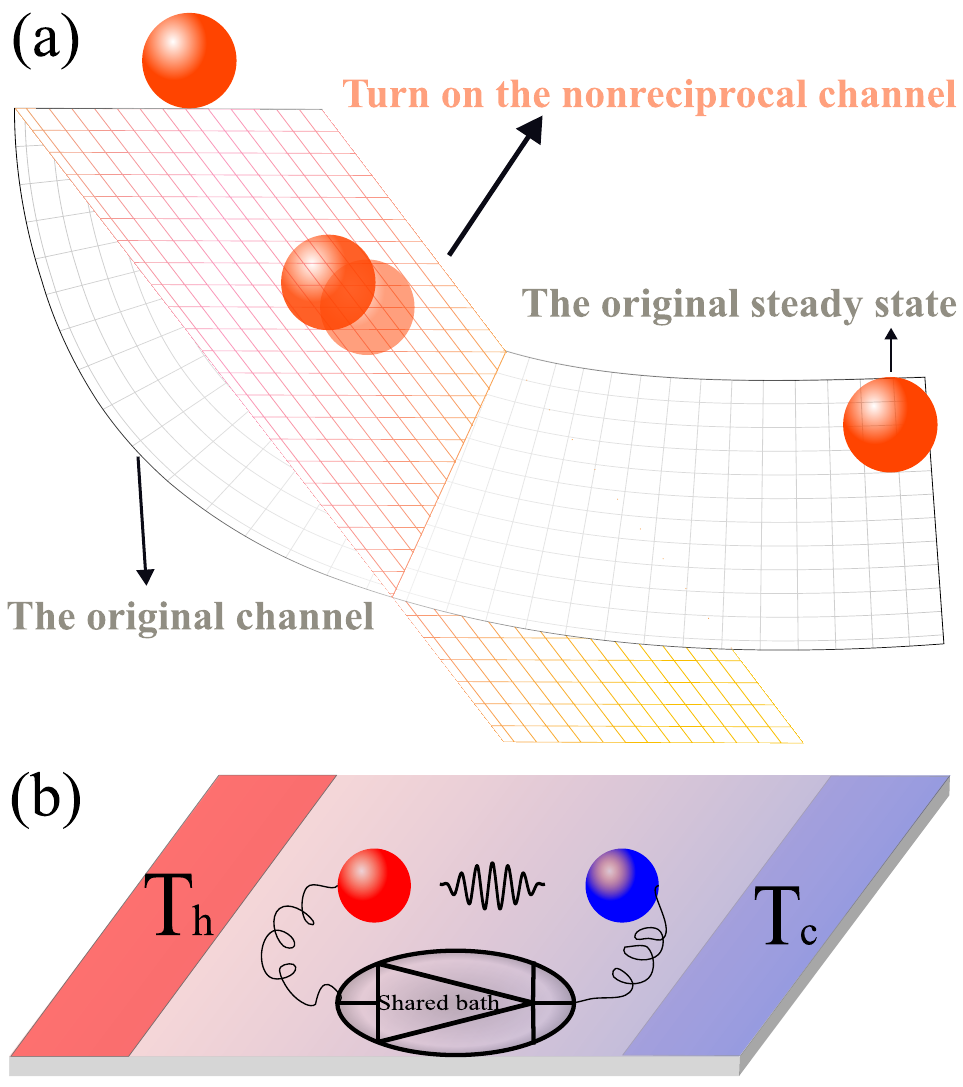}
\caption{(a) Operational principle of nonreciprocal relaxation acceleration. Activating the nonreciprocal coupling for a transient duration significantly expedites the convergence of the quantum state toward the nonequilibrium target steady state. (b) Schematic layout of the model configuration. Two interacting bosonic modes are individually coupled to distinct thermal reservoirs maintained at unequal temperatures. The directional, nonreciprocal interaction between the modes is engineered via a shared common reservoir.}
\label{fig1}
\end{figure}

Assuming resonance condition ($\omega_{1}=\omega_{2}=\omega_{0}$),
we transform into the rotating frame defined by $\rho=e^{i(\omega_{1}a_{1}^{\dagger}a_{1}+\omega_{2}a_{2}^{\dagger}a_{2})t}\tilde{\rho}e^{-i(\omega_{1}a_{1}^{\dagger}a_{1}+\omega_{2}a_{2}^{\dagger}a_{2})t}$,
yielding the Lindblad master equation 
\begin{align}
\frac{d}{dt}\rho= & -i[H,\rho]+\Gamma\mathcal{D}[L]\rho\nonumber \\
 & +\sum_{j=1,2}\kappa_{j}(n_{j}+1)\mathcal{D}[a_{j}]\rho+\kappa_{j}n_{j}\mathcal{D}[a_{j}^{\dagger}]\rho,
\end{align}
where $H=\frac{\lambda}{2}(a_{1}a_{2}^{\dagger}+a_{2}a_{1}^{\dagger})$.
According to Ref. \cite{10.21468/SciPostPhysLectNotes.44}, the evolution of the mean fields $\langle a_{i}\rangle$
reads
\begin{align}
\frac{d}{dt}\langle a_{1}\rangle= & -\frac{\kappa_{1}+\Gamma}{2}\langle a_{1}\rangle-i\frac{\tilde{\lambda}_{12}^{*}}{2}\langle a_{2}\rangle,\nonumber\\
\frac{d}{dt}\langle a_{2}\rangle= & -i\frac{\tilde{\lambda}_{21}}{2}\langle a_{1}\rangle-\frac{\kappa_{2}+\Gamma}{2}\langle a_{2}\rangle,
\label{first_moment_evo}
\end{align}
with $\tilde{\lambda}_{12}\equiv\lambda-\Gamma e^{-i\theta}$ and
$\tilde{\lambda}_{21}\equiv\lambda+\Gamma e^{-i\theta}$. Under the
specific condition $\theta=\pi$ and $\Gamma=\lambda$, we obtain
$\tilde{\lambda}_{12}=2\lambda$ and $\tilde{\lambda}_{21}=0$, which
corresponds to a purely unidirectional influence from the second mode
to the first. Choosing $\theta=0$ under the same coupling strength
($\Gamma=\lambda$) reverses the direction of this nonreciprocity. Setting $\Gamma=0$ recovers the standard reciprocal dissipative dynamics. 

The central paradigm of our approach is outlined in Fig.~\ref{fig1}. Specifically, Fig.~\ref{fig1}(a) demonstrates how a finite-duration nonreciprocal pulse induces a rapid dynamical shortcut to the NESS. Crucially, precise timing of this pulse truncation is imperative when finite-temperature baths cause the nonreciprocal and reciprocal steady states to diverge. Finally, Fig.~\ref{fig1}(b) illustrates the underlying physical configuration, where chiral nonreciprocity is mediated entirely via a shared bath.

\subsection{Mean-Field Dynamics of the Nonreciprocal Relaxation}
\begin{figure}
\includegraphics[scale=0.7]{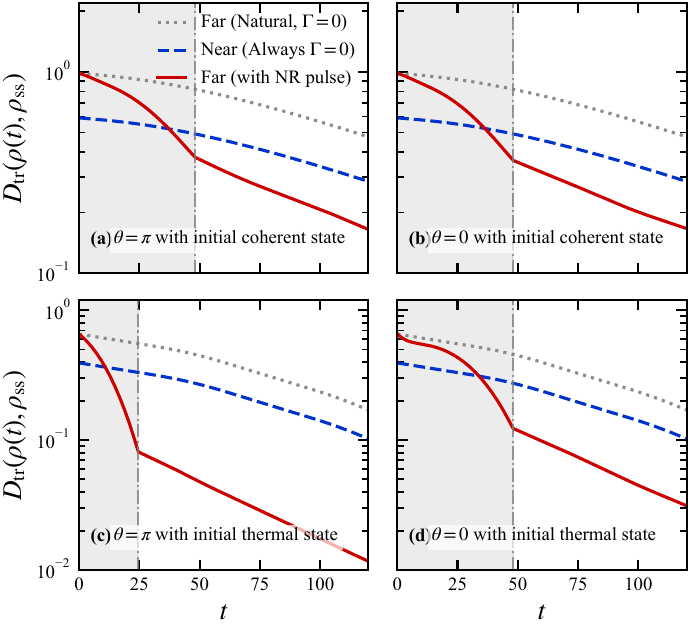}
\caption{Dynamics of nonreciprocal relaxation acceleration, quantified by the time-dependent trace distance between the instantaneous state and the reciprocal NESS $D_{\rm tr}(\rho(t), \rho_{\rm ss})$. A distant initial state, far from the reciprocal NESS, relaxes significantly faster than a proximal ("near") initial state when assisted by a transient activation of the nonreciprocal channel. Specifically, the distant initial states are prepared as a two-mode coherent state $\rho_{\text{far}} = \rho_{\text{coh}}^{(1)}(1.4) \otimes \rho_{\text{coh}}^{(2)}(1.4)$ for (a) and (b) and thermal state $\rho_{\text{far}} = \rho_{\text{th}}^{(1)}(3.0) \otimes \rho_{\text{th}}^{(2)}(0.05)$ for (c) and (d) with $\rho_{\text{th}}^{(i)}(\bar{n})$ denoting a local thermal state of the $i$-th mode with an average photon number $\bar{n}$, while the proximal state is constructed as a statistical mixture $\rho_{\text{near}} = 0.6\rho_{\text{far}} + 0.4\rho_{\text{ss}}$, where $\rho_{\text{ss}}$ is the target reciprocal NESS. The bipartite bosonic Fock space is truncated to a constant local dimension of $N_c = 16$ per mode. Panels (a,c) and (b,d) correspond to the phase conditions $\theta=\pi$ and $\theta=0$, which dictate unidirectional (chiral) coupling from the second mode to the first, and from the first mode to the second, respectively. The system parameters are chosen as follows: coherent inter-mode coupling $\lambda=0.1$, local dissipation rates $\kappa_{1}=\kappa_{2}=0.02$, and reservoir thermal occupations $n_1=0.15$ and $n_2=0.05$ (implying a non-zero temperature gradient). The nonreciprocal dissipation strength is set to $\Gamma=\lambda$ to satisfy the perfect nonreciprocity condition. The optimal duration of the nonreciprocal pulse is determined by truncating the channel at the instant where the transient state minimizes its trace distance to the original NESS, beyond which the distance would otherwise rebound and monotonically increase.}
\label{fig2}
\end{figure}

To elucidate the explicit impact of nonreciprocal dissipation on the characteristic relaxation timescales, we represent the mean-field evolution of the expectations $\langle a_1 \rangle$ and $\langle a_2 \rangle$ in the standard dynamical matrix form
\begin{align}
\frac{d}{dt}\left(\begin{array}{c}
\langle a_{1}\rangle\\
\langle a_{2}\rangle
\end{array}\right)= & \left(\begin{array}{cc}
-\frac{\kappa_{1}+\Gamma}{2} & -i\frac{\tilde{\lambda}_{12}^{*}}{2}\\
-i\frac{\tilde{\lambda}_{21}}{2} & -\frac{\kappa_{2}+\Gamma}{2}
\end{array}\right)\left(\begin{array}{c}
\langle a_{1}\rangle\\
\langle a_{2}\rangle
\end{array}\right)=U\left(\begin{array}{c}
\langle a_{1}\rangle\\
\langle a_{2}\rangle
\end{array}\right).\label{first_Evo}
\end{align} 
Under the symmetric dissipation condition $\kappa_{1}=\kappa_{2}=\kappa$, the eigenvalues of the evolution matrix $U$ take the form $\mu_{\pm}=-\frac{\kappa+\Gamma}{2}\pm\frac{1}{2}\sqrt{-\tilde{\lambda}_{12}^{*}\tilde{\lambda}_{21}}$. In the absence of the nonreciprocal channel ($\Gamma=0$), the system exhibits standard reciprocal decay with $\mu_{\pm}=-\frac{\kappa}{2}\pm i\frac{\lambda}{2}$. Conversely, when the nonreciprocity is perfectly tuned ($\Gamma e^{-i\theta}=-\lambda$ or $\Gamma e^{i\theta}=\lambda$), the eigenvalues become strictly degenerate, yielding $\mu_{\pm}=-\frac{\kappa+\lambda}{2}$. Because the real part of the evolution matrix's eigenvalues fundamentally determines the overall relaxation rate, this mathematical transition demonstrates how nonreciprocity leverages the inter-mode coupling strength $\lambda$ to induce a macroscopic relaxation speedup, a phenomenon clearly visualized in Fig.~\ref{fig2}. Here, the vertical axis tracks the state's convergence using the trace distance, rigorously defined as \citep{nielsen_quantum_2010} $D_{\text{tr}}(\rho(t), \rho_{\text{ss}}) = \frac{1}{2} \text{Tr}\sqrt{(\rho(t) - \rho_{\text{ss}})^\dagger (\rho(t) - \rho_{\text{ss}})}$. As plotted on the logarithmic scale, the conspicuously steep descent of the solid red curves within the shaded pulse regions provides direct macroscopic evidence of this nonreciprocal acceleration against the natural decay background.

A remarkable feature of the nonreciprocal accelerated relaxation is that it is fundamentally invariant under the reversal of the nonreciprocal direction, a property mathematically evidenced by the spectral analysis above. Although the phase condition $\Gamma e^{-i\theta}=-\lambda$ dictates a unidirectional coupling from the second mode to the first, and $\Gamma e^{i\theta}=\lambda$ strictly reverses this chirality, both configurations yield an identical accelerated relaxation rate governed by the degenerate eigenvalue $\mu_{\pm}=-\frac{\kappa+\lambda}{2}$. Consequently, the dynamical speedup persists regardless of whether the macroscopic steady-state heat flux aligns with or opposes the engineered chiral direction.

Notably, the above explanation based on $\langle a_i\rangle$ successfully describes general initial states, such as coherent states, but is inapplicable to thermal states, for which $\langle a_i\rangle$ identically vanishes. Besides, this kinetic speedup is optimally observed in the zero-temperature vacuum limit under asymmetric local dissipation, because the two-mode vacuum acts as a common dark state unperturbed by the nonreciprocal activation. While rising ambient temperatures introduce thermal fluctuations that inevitably obscure this acceleration (thereby imposing strict thermal bounds on experimental implementations), the mean-field first moments are fundamentally blind to this temperature dependence. Uncovering the complete physical picture of these phenomena therefore requires a rigorous treatment of the system's quantum fluctuations via the second-order covariance matrix.

\subsection{Quantum Fluctuations and State Fidelity}

To achieve a comprehensive understanding of the nonreciprocal relaxation speedup and its associated quantum and thermal fluctuations, we transition to a phase-space representation. We introduce the standard position and momentum quadratures $x_{j}=\frac{a_{j}+a_{j}^{\dagger}}{\sqrt{2}}$ and $p_{j}=\frac{a_{j}-a_{j}^{\dagger}}{i\sqrt{2}}$ ($j=1,2$), grouping them into the quadrature operator vector $\boldsymbol{R}=(x_{1},p_{1},x_{2},p_{2})^{T}$. The fluctuation profile of the system is then entirely encapsulated within the covariance matrix $\boldsymbol{V}$, whose elements are defined as $V_{ij}=\frac{1}{2}\langle\{R_{i},R_{j}\}\rangle-\langle R_{i}\rangle\langle R_{j}\rangle$. Following Eq. (\ref{first_Evo}), the mean-field evolution of these quadratures is governed by the linear dynamical equation $\frac{d}{dt}\langle\boldsymbol{R}\rangle=A\langle\boldsymbol{R}\rangle$, where the drift matrix $A$ takes the explicit form
\begin{align}
A & =\left(\begin{array}{cccc}
-\frac{\kappa_{1}+\Gamma}{2} & 0 & -\frac{1}{2}\Gamma\sin\theta & \frac{\lambda-\Gamma\cos\theta}{2}\\
0 & -\frac{\kappa_{1}+\Gamma}{2} & -\frac{\lambda-\Gamma\cos\theta}{2} & -\frac{1}{2}\Gamma\sin\theta\\
-\frac{1}{2}\Gamma\sin\theta & \frac{\lambda+\Gamma\cos\theta}{2} & -\frac{\kappa_{2}+\Gamma}{2} & 0\\
-\frac{\lambda+\Gamma\cos\theta}{2} & -\frac{1}{2}\Gamma\sin\theta & 0 & -\frac{\kappa_{2}+\Gamma}{2}
\end{array}\right).\label{drift}
\end{align}
where the perfectly nonreciprocal conditions ($\Gamma=\lambda$ with $\theta=0$ or $\theta=\pi$) reduce the drift matrix to a lower or upper triangular form, respectively, explicitly dictating the directionality of the nonreciprocal transport.

To evaluate the time evolution of the covariance matrix $V$, we must first determine the diffusion matrix $D$, which dictates the noise injection dynamics via the differential Lyapunov equation $\frac{d}{dt}V=AV+VA^{T}+D$. The elements of this diffusion matrix are explicitly formulated as
\begin{equation}
D_{mn}=\sum_{k}\frac{\gamma_{k}}{2}\langle[L_{k}^{\dagger},R_{m}][R_{n},L_{k}]+[L_{k}^{\dagger},R_{n}][R_{m},L_{k}]\rangle
\end{equation}
from the Appendix \ref{appendixA}, where $L_{k}$ denotes the $k$-th dissipation jump operator with the corresponding decay rate $\gamma_{k}$. Notably, this definition underscores that the diffusion matrix originates purely from irreversible environmental dissipation, remaining strictly independent of the coherent unitary evolution driven by the Hamiltonian. By expanding the Lindblad operators in the quadrature basis as $L_{k}=\boldsymbol{c}_{k}^{T}\boldsymbol{R}=\sum_{j}c_{k}^{j}R^{j}$ with complex coefficient vectors $\boldsymbol{c}_{k}=\boldsymbol{u}_{k}+i\boldsymbol{v}_{k}$, one can analytically reduce the diffusion matrix to $D=\sum_{k}\gamma_{k}(\boldsymbol{u}_{k}\boldsymbol{u}_{k}^{T}+\boldsymbol{v}_{k}\boldsymbol{v}_{k}^{T})$. Applying this formalism to our specific nonreciprocal dissipators, we arrive at the exact analytical expression for the diffusion matrix
\begin{equation}
D=\left(\begin{array}{cccc}
\xi_{1} & 0 & \frac{\Gamma}{2}\sin\theta & \frac{\Gamma}{2}\cos\theta\\
0 & \xi_{1} & -\frac{\Gamma}{2}\cos\theta & \frac{\Gamma}{2}\sin\theta\\
\frac{\Gamma}{2}\sin\theta & -\frac{\Gamma}{2}\cos\theta & \xi_{2} & 0\\
\frac{\Gamma}{2}\cos\theta & \frac{\Gamma}{2}\sin\theta & 0 & \xi_{2}
\end{array}\right),
\end{equation}
with the diagonal noise terms $\xi_{i}\equiv\frac{\Gamma}{2}+\kappa_{i}(n_{i}+\frac{1}{2})$.
Hereafter, imposing symmetric dissipation $\kappa_{1}=\kappa_{2}=\kappa$, we can therefore solve for the steady-state covariance matrix via the algebraic Lyapunov equation $AV^{\text{ss}}+V^{\text{ss}}A^{T}+D=0$.

To gain a deeper insight into the mechanism of nonreciprocal relaxation acceleration, we explicitly construct the drift and diffusion matrices, $A_{G}$ and $D_{G}$, which characterize the evolution toward the nonreciprocal steady state $V_{G}^{\mathrm{ss}}$. The transient covariance matrix can be analytically expressed as
\begin{equation}
	V_{G}(t)=V_{G}^{\mathrm{ss}}+e^{A_{G}t}(V(0)-V_{G}^{\mathrm{ss}})e^{A_{G}^{T}t}.\label{transient}
\end{equation} 
In the absence of nonreciprocal coupling ($\Gamma=0$), the intrinsic drift and diffusion matrices, $A_{0}$ and $D_{0}$, determine the target reciprocal steady state $V_{0}^{\mathrm{ss}}$. Under these reciprocal conditions, the corresponding transient covariance matrix satisfies an analogous relation to Eq. (\ref{transient}).

According to Eq. (\ref{drift}), the evolution of the reciprocal transient covariance matrix is governed by
\begin{equation}
	||V_0(t)-V_0^{\text{ss}}||_F=e^{-\kappa t}||V(0)-V_0^{\text{ss}}||_F,\label{reF}
\end{equation}
where $||\cdot||_F$ denotes the Frobenius norm.
Applying a similar procedure to the nonreciprocal case yields
\begin{align}
	V_G(t) - V_G^{\mathrm{ss}} ={} & e^{-(\kappa+\lambda)t} \left[ \mathbf{I} + t \left( A_G + \frac{\kappa + \lambda}{2} \mathbf{I} \right) \right] \nonumber\\
	& \times \left[ V(0) - V_G^{\mathrm{ss}} \right] \nonumber\\
	& \times \left[ \mathbf{I} + t \left( A_G^T + \frac{\kappa + \lambda}{2} \mathbf{I} \right) \right],
\end{align}
whose Frobenius norm is bounded by
\begin{equation}
	||V_G(t) - V_G^{\mathrm{ss}}||_F \le e^{-(\kappa+\lambda)t}(1+\lambda t)^2||V(0)-V_G^{\text{ss}}||_F.\label{nonreF}
\end{equation}
These two cases yield asymptotic relaxation rates, defined by $-\lim_{t\rightarrow\infty}\frac{1}{t}\ln||V(t)-V^{\text{ss}}||_F$, of $\kappa$ and at least $\kappa+\lambda$ for the reciprocal and nonreciprocal dynamics, respectively. This explicitly demonstrates the nonreciprocal relaxation acceleration from the perspective of the covariance matrix, regardless of whether the system starts from a thermal state.

By applying the triangle inequality along with Eqs. (\ref{reF}) and (\ref{nonreF}), we obtain
\begin{align}
	\| V_G(t) - V_0^{\mathrm{ss}} \|_F \le{} & \| V_G^{\mathrm{ss}} - V_0^{\mathrm{ss}} \|_F \nonumber\\
	& + e^{-(\kappa+\lambda)t} (1+ \lambda t)^2 \| V(0) - V_G^{\mathrm{ss}} \|_F.
\end{align}
Consequently, the nonreciprocal acceleration is achieved, i.e., $||V_G(t)-V_0^{\text{ss}}||_F\le||V_0(t)-V_0^{\text{ss}}||_F$, provided that
\begin{align}
	& \| V_G^{\mathrm{ss}} - V_0^{\mathrm{ss}} \|_F  + e^{-(\kappa+\lambda)t} (1+ \lambda t)^2 \| V(0) - V_G^{\mathrm{ss}} \|_F \nonumber\\
	\le&  e^{-\kappa t} \| V(0) - V_0^{\mathrm{ss}} \|_F.\label{sufficient}
\end{align}
Equation (\ref{sufficient}) thus serves as a sufficient condition for realizing nonreciprocal relaxation acceleration. Notably, at higher bath temperatures, the first term on the left-hand side of Eq. (\ref{sufficient}) increases, rendering the nonreciprocal relaxation acceleration more difficult to achieve.

\section{Instantaneous current}\label{sec3}

In this section, we investigate the phenomenon of nonreciprocal relaxation acceleration from a thermodynamic perspective by analyzing the transient heat currents. Assuming resonant bosonic modes with $\omega_{1}=\omega_{2}=\omega_{0}=1$ in the original Hamiltonian $H_0$ [Eq. (\ref{originalH})], we define the heat current flowing from the $j$th thermal bath into the system as $J_{j}=\omega_{0}\kappa_j(n_{j}-\langle a^{\dagger}_j a_j\rangle)$ \citep{PRXQuantum.5.020201}.

\begin{figure}
\includegraphics[scale=0.7]{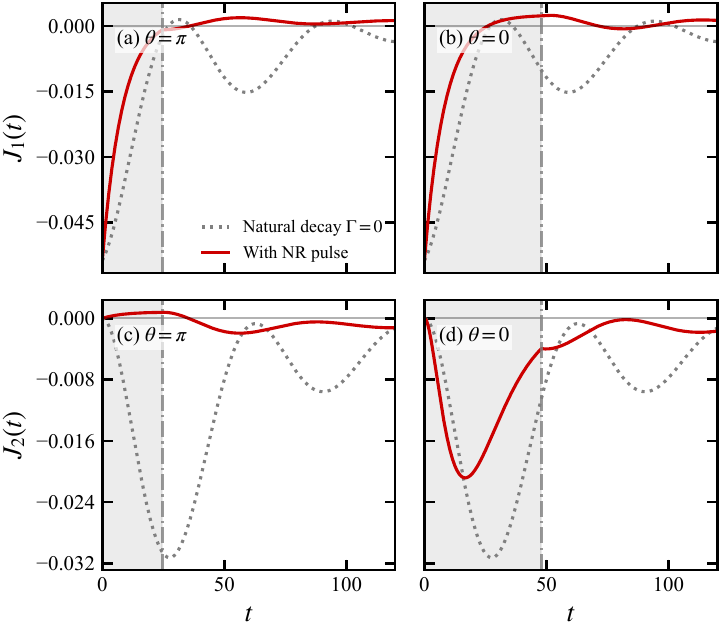}
\caption{Instantaneous heat currents $J_1(t)$ and $J_2(t)$ exchanged with the local reservoirs, flowing from the respective reservoirs into the system. The left column, (a) and (c), corresponds to the unidirectional coupling phase $\theta=\pi$, whereas the right column, (b) and (d), corresponds to $\theta=0$. All panels highlight the relaxation from a highly excited initial state, comparing natural reciprocal decay ($\Gamma=0$, gray dotted curves) with the nonreciprocal (NR) pulse-assisted evolution (solid curves). During the pulse duration (shaded regions), the engineered nonreciprocity dynamically redirects the inter-mode energy transport. By preventing the thermal energy from oscillating back and forth between the two modes, the nonreciprocal coupling forces the excess heat to drain rapidly into a specific reservoir. As a result, the thermodynamic fluxes undergo a significantly accelerated convergence toward the NESS.}
\label{fig3}
\end{figure}

Figure~\ref{fig3} depicts the temporal evolution of these heat currents under the nonreciprocal phase conditions $\theta=\pi$ and $\theta=0$. Here, we employ the same farther initial thermal state analyzed in Fig.~\ref{fig2}, as thermal states are of central importance in the quantum thermodynamic regime. Figures \ref{fig3}(a) and \ref{fig3}(c) depict the scenario of unidirectional coupling directed from the second mode to the first ($\theta=\pi$), whereas Figs. \ref{fig3}(b) and \ref{fig3}(d) represent the reverse direction ($\theta=0$). In the former case, the nonreciprocal interaction significantly suppresses the pronounced thermal oscillations that characterize natural reciprocal decay. Initially, because the energy transport from the first mode to the second is nonreciprocally blocked, the substantial heat exchange between the first mode and its local bath exerts a negligible influence on the second mode. By contrast, under natural reciprocal decay, the heat current of the second mode exhibits high-amplitude oscillations driven by continuous inter-mode energy sloshing. Once the nonreciprocal channel is deactivated, the heat currents in the nonreciprocal scheme remain minimal, having rapidly approached their steady-state values, whereas the natural decay continues to manifest large transient macroscopic currents. Consequently, the nonreciprocal protocol enables the thermodynamic fluxes to stabilize on a remarkably shorter timescale. For the reverse configuration ($\theta=0$), the initial heat current of the second mode exhibits a larger transient oscillation than in the $\theta=\pi$ case, a direct consequence of the reversed unidirectional energy flow funneling excess excitations into the second mode.

These thermodynamic analyses underscore the necessity of optimizing the activation duration of the nonreciprocal channel. The pulse truncation instants numerically identified in Fig.~\ref{fig2} closely coincide with the moments when both macroscopic heat currents have been quenched to near-zero values in Fig.~\ref{fig3}. Terminating the nonreciprocal pulse at this optimal juncture prevents subsequent thermodynamic oscillations, thereby locking the system into a trajectory of accelerated convergence toward the NESS.

\section{Physical Implementation via Temporal Pulse Control}\label{sec4}
To physically realize the transient activation of the nonreciprocal channel, we introduce a time-dependent driving scheme. The composite density matrix of the two-mode system and the engineered shared bath evolves according to
\begin{align}
\frac{d}{dt}\tilde{\rho}_{S+B}= & -i[H_{\mathrm{lab}}(t),\tilde{\rho}_{S+B}]\nonumber \\
 & +\!\!\sum_{j=1,2}\kappa_{j}\Big((n_{j}\!+\!1)\mathcal{D}[a_{j}]\tilde{\rho}_{S+B}\!+\! n_{j}\mathcal{D}[a_{j}^{\dagger}]\tilde{\rho}_{S+B}\Big),
\end{align}
where $H_{\mathrm{lab}}(t)=H_{\mathrm{sys}}+H_{\mathrm{bath}}+H_{\mathrm{int}}(t)$ denotes the total Hamiltonian in the laboratory frame. 
The unperturbed dynamics are dictated by the system Hamiltonian $H_{\mathrm{sys}}=\omega_{0}a_{1}^{\dagger}a_{1}+\omega_{0}a_{2}^{\dagger}a_{2}+\frac{\lambda}{2}(a_{1}^{\dagger}a_{2}+a_{1}a_{2}^{\dagger})$ and the bath Hamiltonian $H_{\mathrm{bath}}=\sum_{k}\omega_{k}b_{k}^{\dagger}b_{k}$. 

To explicitly engineer the desired nonreciprocity, the system is subjected to an external temporal pulse, leading to the interaction Hamiltonian
\begin{align}
H_{\mathrm{int}}(t)& =\sum_{j=1,2}\eta(t)\cos(\omega_{p}t+\phi_{j})(a_{j}+a_{j}^{\dagger})\sum_{k}g_{k}(b_{k}+b_{k}^{\dagger}).
\end{align}
The nonreciprocal behavior is engineered by tailoring the controlled driving pulse $\eta(t)\cos(\omega_{p}t+\phi_{j})$. Moving to the interaction picture defined by the free-evolution operator $U_{0}(t)=\exp\Big[-i\Big(\omega_{0}a_{1}^{\dagger}a_{1}+\omega_{0}a_{2}^{\dagger}a_{2}+H_{\mathrm{bath}}\Big)t\Big]$, the quantum master equation transforms into
\begin{align}
\frac{d}{dt}\rho_{S+B}= & -i\Big[\frac{\lambda}{2}(a_{1}^{\dagger}a_{2}+a_{1}a_{2}^{\dagger})+\tilde{H}_{\mathrm{int}}(t),\rho_{S+B}\Big]\nonumber \\
 & +\sum_{j=1,2}\kappa_{j}\Big((n_{j}+1)\mathcal{D}[a_{j}]\rho_{S+B}+n_{j}\mathcal{D}[a_{j}^{\dagger}]\rho_{S+B}\Big).
\end{align}
The terms in the interaction Hamiltonian $\tilde{H}_{\mathrm{int}}(t)$ take the form
\begin{align}
\frac{1}{2}e^{i(\omega_{p}t+\phi_{j})}a_{j}e^{-i\omega_{0}t}b_{k}^{\dagger}e^{i\omega_{k}t} & =\frac{1}{2}e^{i\phi_{j}}a_{j}b_{k}^{\dagger}e^{i(\omega_{p}-\omega_{0}+\omega_{k})t}.
\end{align}
By imposing the rotating-wave approximation (RWA) to neglect rapidly oscillating counter-rotating terms, the effective interaction Hamiltonian simplifies to
\begin{align}
\tilde{H}_{\mathrm{int}}^{\mathrm{RWA}}(t) & =\sum_{k}\frac{\eta(t)g_{k}}{2}\Big[e^{i\phi_{1}}a_{1}b_{k}^{\dagger}+e^{i\phi_{2}}a_{2}b_{k}^{\dagger}+\mathrm{h.c.}\Big]\nonumber \\
 & =\sum_{k}\frac{\eta(t)g_{k}}{2}\Big[e^{i\phi_{1}}(a_{1}-ie^{i\theta}a_{2})b_{k}^{\dagger}+\mathrm{h.c.}\Big],
\end{align}
where $\theta = \pi/2 + \phi_{2} - \phi_{1}$ denotes the relative phase of the driving fields.

Assuming the engineered shared bath is maintained at zero temperature, tracing out its degrees of freedom under the standard Born-Markov approximation yields a time-dependent collective dissipation rate $\Gamma(t)\propto|\eta(t)|^{2}$. Consequently, the effective quantum master equation governing the reduced density matrix of the system, $\rho_{S}$, is rigorously cast into the form 
\begin{align}
\frac{d}{dt}\rho_{S}= & -i[\frac{\lambda}{2}(a_{1}^{\dagger}a_{2}+a_{1}a_{2}^{\dagger}),\rho_{S}]+\Gamma(t)\mathcal{D}[a_{1}-ie^{i\theta}a_{2}]\rho_{S}\nonumber \\
 & +\sum_{j=1,2}\kappa_{j}\Big((n_{j}+1)\mathcal{D}[a_{j}]+n_{j}\mathcal{D}[a_{j}^{\dagger}]\Big)\rho_{S}.
\end{align}
This expression elegantly encapsulates the physical blueprint of our nonreciprocal acceleration scheme. By precisely modulating the temporal envelope $\eta(t)$ of the external driving field, one can dynamically toggle the collective nonreciprocal dissipation channel on and off. Concurrently, the relative driving phase $\theta$ dictates the chiral direction of the inter-mode transport. Therefore, applying a tailored, finite-duration pulse temporarily forces the system along a highly accelerated nonreciprocal trajectory, successfully facilitating a rapid shortcut to the target nonequilibrium steady state once the pulse is extinguished.

\section{conclusions and outlook}\label{sec5}
We have investigated a two-mode bosonic model coupled to distinct local thermal baths to demonstrate the phenomenon of nonreciprocal relaxation acceleration. We reveal that the transient activation of a nonreciprocal coupling channel significantly expedites the thermalization process. Crucially, the optimal duration of this nonreciprocal pulse closely corresponds to the dynamical regime where both macroscopic heat currents are effectively suppressed. By analyzing the temperature dependence of this effect, we show that the relaxation speedup is maximally pronounced in the low-temperature limit, highlighting its direct relevance to modern open quantum system experiments. We further rigorously establish that the emergence of this acceleration is fundamentally independent of the direction of the nonreciprocity. Future research directions will focus on generalizing this nonreciprocal speedup to many-body systems and optimizing relaxation protocols through extended non-Hermitian dynamics.

We note that during the finalization of this manuscript, a related yet distinct work by Yan et al. \cite{yan2026nonreciprocal} appeared, which proposed a nonreciprocal quantum Mpemba effect based on reservoir-parameter swapping and Liouvillian eigenvector rotation.

\begin{acknowledgments}
	This work was financially supported from the National Natural Science Foundation of China (Grants Nos. 12475039, 12075199, 11704093, 12347151, 12247172 and 12505032) and Guangdong Basic and Applied Basic Research Foundation (No. 2025A1515010350). 
\end{acknowledgments}

\appendix

\section{Derivation of the Diffusion Matrix}\label{appendixA}

For our Gaussian systems, to get the covariance matrix, we first need
to obtain the drift matrix $A$ and diffusion matrix $D$ in the Lyapunov
equation. The drift matrix $A$ is easy to obtain by the first order
derivative of $\langle\boldsymbol{R}\rangle$. To obtain the diffusion
matrix $D$, we first note that the diffusion matrix is only related
to the dissipation channels linearly rather than the unitary evolution.
Therefore, we focus on one dissipation channel with the jump operator,
which can be written as $L=\boldsymbol{c}^T\boldsymbol{R}=\boldsymbol{u}^{T}\boldsymbol{R}+i\boldsymbol{v}^{T}\boldsymbol{R}$
with the quantum master equation $\frac{d}{dt}\rho=\gamma\mathcal{D}[L]\rho=\gamma(L\rho L^{\dagger}-\frac{1}{2}\{L^{\dagger}L,\rho\})$.

According to the adjoint master equation for $\frac{d}{dt}\langle O\rangle=\gamma\langle\mathcal{D}^{*}(O)\rangle=\gamma(L^{\dag}OL-\frac{1}{2}\{L^{\dagger}L,O\})$,
\begin{widetext}
\begin{align}
\dot{V}_{mn}= & \frac{\gamma}{2}\langle\mathcal{D}^{*}(R_{m}R_{n}+R_{n}R_{m})\rangle-\langle\dot{R}_{m}\rangle\langle R_{n}\rangle-\langle R_{m}\rangle\langle\dot{R}_{n}\rangle\nonumber\\
= & \frac{\gamma}{2}\langle\mathcal{D}^{*}(R_{m})R_{n}\!\!+\!\! R_{m}\mathcal{D}^{*}(R_{n})\!\!+\!\!\mathcal{D}^{*}(R_{n})R_{m}\!\!+\!\! R_{n}\mathcal{D}^{*}(R_{m})\rangle \!\!+\!\!\frac{\gamma}{2}\langle[L^{\dagger},R_{m}][R_{n},L]\!\!+\!\![L^{\dagger},R_{n}][R_{m},L]\rangle \!\!-\!\!\langle\dot{R}_{m}\rangle\langle R_{n}\rangle\!\!-\!\!\langle R_{m}\rangle\langle\dot{R}_{n}\rangle\nonumber\\
= & \frac{1}{2}\langle\sum_{k} A_{mk}R_{k}R_{n}+\sum_{k} A_{nk}R_{m}R_{k}+\sum_{k} A_{nk}R_{k}R_{m}+\sum_{k} A_{mk}R_{n}R_{k}\rangle\nonumber\\
&-\langle\sum_{k}A_{mk}R_{k}\rangle\langle R_{n}\rangle-\langle R_{m}\rangle\langle\sum_{k}A_{nk}R_{k}\rangle+\frac{\gamma}{2}\langle[L^{\dagger},R_{m}][R_{n},L]+[L^{\dagger},R_{n}][R_{m},L]\rangle\nonumber\\
= & \sum_{k}A_{mk}V_{kn}+\sum_{k} V_{mk}A_{nk}+\frac{\gamma}{2}\langle[L^{\dagger},R_{m}][R_{n},L]+[L^{\dagger},R_{n}][R_{m},L]\rangle\nonumber\\
= & (AV+VA^{T})_{mn}+\frac{\gamma}{2}\langle[L^{\dagger},R_{m}][R_{n},L]+[L^{\dagger},R_{n}][R_{m},L]\rangle
\end{align}
\end{widetext}
Therefore, the diffusion matrix can be expressed as
\begin{equation}
	D_{mn}=\frac{\gamma}{2}\langle[L^{\dagger},R_{m}][R_{n},L]+[L^{\dagger},R_{n}][R_{m},L]\rangle.\label{Dmn}
\end{equation}
With the decomposition of $L=\boldsymbol{c}^T\boldsymbol{R}=\boldsymbol{u}^{T}\boldsymbol{R}+i\boldsymbol{v}^{T}\boldsymbol{R}$ and $[R_m, R_n]=i\Omega_{mn}$, we can obtain
\begin{align}
	[R_n,L]&=i(\Omega \boldsymbol{c})_n\nonumber\\
	[L^\dagger,R_m]&=-i(\Omega \boldsymbol{c}^{*})_m.
\end{align}
Then substitute these two equations into the Eq. (\ref{Dmn}), we get $D=\gamma\Omega(\boldsymbol{u}\boldsymbol{u}^T+\boldsymbol{v}\boldsymbol{v}^T)\Omega^T$ with the sympletic form $\Omega=\bigoplus_{i=1}^{2} \begin{pmatrix} 0 & 1 \\ -1 & 0 \end{pmatrix}$. When the jump operator depends only on the annihilation operators or creation operators, the symplectic transformation only exchange the two matrices $\boldsymbol{u}\boldsymbol{u}^T\leftrightarrow\boldsymbol{v}\boldsymbol{v}^T$. Hence, the final expression of the diffusion matrix is $D=\gamma(\boldsymbol{u}\boldsymbol{u}^T+\boldsymbol{v}\boldsymbol{v}^T)$.

%\bibliography{reference}

\end{document}